\documentstyle[new_cite,epsf]{mn}

\begin{document}

\title[Constraints on a soft excess in 3C 279]{Constraints on a soft X-ray excess in the quasar 3C 279}
\author[A.J. Lawson and I.M. M$^{\rm
c}$Hardy]{A.~J.~Lawson and I.~M.~M$^{\rm c}$Hardy\\ Department of
Physics \& Astronomy, University of Southampton, University Road,
Southampton SO17 1BJ} 
\date{Accepted for publication by MNRAS, 1998 July 3.}
\maketitle
\begin{abstract}
We present the results of a three-week daily monitoring campaign on
the quasar 3C 279 by the X-ray satellites {\sl RXTE} and {\sl
ROSAT}. A cross correlation provides no evidence for any time lag
between the very similar soft and hard X-ray light curves, and the
source shows no significant spectral variability over the observing
period. There is no evidence to support the presence of a soft excess,
with a 99 per cent upper limit on any such component of 25 per cent of
the total observed luminosity in the 0.1--2\,keV band
($<\!3\times10^{38}$\,W). This fraction (but not the luminosity) is
significantly less than that of the soft excess observed in 3C 273.
\end{abstract}

\section{Introduction}
Soft excesses are the steep upturn in the X-ray spectra of active
galactic nuclei (AGN) below $\sim$2\,keV (e.g. \ncite{MUS93}). They
were first detected in Seyfert galaxies \cite{ARN85}, where they are
quite common (\ncite{TAP89}; Mushotzky et al. 1993). They can be
modelled by a steep power law ($\alpha\!>$2), or by a cool thermal
component (T$<\!150$\,eV), and are thought to be the high energy tail
of the accretion disc radiation (e.g. Mushotzky et al. 1993).

Soft excesses are reasonably common in quasars
\cite{URR89,MAS92,SAX93}, and there appears to be a trend for the more
distant objects to have flatter spectra \cite{SCH92}, as expected if a
soft component were being redshifted below the soft X-ray energy
range. Initial detections were in radio-quiet objects \cite{COM92}, but
more recent work \cite{BUH95,PRI96,SCH96} suggests that a soft excess
may also be a common feature in radio-loud objects. One of these with
a consistently detected soft excess is 3C 273
\cite{TUR90,STA92B,LEA95}. It is also one object for which the
spectral parameters of the soft excess itself are reasonably well
known (Leach et al. 1995).

\section{The {\sl RXTE} and {\sl ROSAT} Observations}
\label{sec-obs}

\begin{table*}
\begin{minipage}{171 mm}
\caption{Observational parameters}
\label{tab-xteobs}
\begin{tabular}{@{}llcrccclcccc@{}}\hline
\multicolumn{6}{c}{\sl RXTE} & & \multicolumn{5}{c}{\sl ROSAT}\\
Obs$^a$ &  Day and time$^b$  & Date$^c$ & Exp$^d$
& Det$^e$ & Count rate$^f$ &  &Obs$^a$ & Day and Time$^b$ &      Date$^c$ & Exp$^d$ & Count rate\\\hline
1  & 27/6 07:01:21 & 179.293 & 1152 & 5 & 6.25$\pm$0.17 & &   1 & 28/6  08:59:35 & 180.375 & 1975  & 0.253$\pm$0.015\\       
2  & 28/6 16:02:41 & 180.669 & 640  & 5 & 5.94$\pm$0.21 & &   2 & 29/6  11:57:16 & 181.498 & 2387  & 0.264$\pm$0.014\\       
3  & 29/6 11:16:09 & 181.470 & 704  & 3 & 5.94$\pm$0.25 & &   3 & 30/6  07:09:45 & 182.298 & 2177  & 0.269$\pm$0.014\\       
4  & 30/6 05:23:29 & 182.225 & 544  & 3 & 5.83$\pm$0.30 & &   4 & 01/7  07:02:43 & 183.294 & 2070  & 0.307$\pm$0.015\\       
5  & 01/7 06:29:13 & 183.270 & 720  & 3 & 5.82$\pm$0.25 & &   5 & 02/7  06:55:30 & 184.289 & 2057  & 0.265$\pm$0.015\\       
6  & 02/7 08:06:09 & 184.338 & 704  & 5 & 6.31$\pm$0.20 & &   6 & 03/7  06:46:37 & 185.282 & 2113  & 0.267$\pm$0.014\\       
7  & 03/7 12:55:21 & 185.538 & 688  & 5 & 5.13$\pm$0.20 & &   7 & 04/7  06:38:45 & 186.277 & 2093  & 0.267$\pm$0.014\\       
8  & 04/7 11:20:09 & 186.472 & 720  & 5 & 5.02$\pm$0.20 & &   8 & 05/7  06:29:45 & 187.271 & 1977  & 0.253$\pm$0.014\\       
9  & 05/7 06:33:37 & 187.273 & 544  & 3 & 5.48$\pm$0.29 & &   9 & 06/7  07:54:21 & 188.329 & 2311  & 0.243$\pm$0.014\\       
10 & 07/7 13:20:25 & 189.556 & 560  & 3 & 4.55$\pm$0.29 & &   10 & 07/7  07:46:29 & 189.324 & 2231  & 0.220$\pm$0.014\\      
11 & 09/7 16:45:53 & 191.699 & 736  & 3 & 4.36$\pm$0.27 & &   11 & 08/7  04:28:53 & 190.187 & 2033  & 0.209$\pm$0.014\\      
12 & 10/7 18:28:25 & 192.770 & 688  & 3 & 4.70$\pm$0.29 & &   12 & 09/7  01:06:24 & 191.046 & 1768  & 0.238$\pm$0.015\\      
13 & 11/7 19:25:21 & 193.810 & 544  & 5 & 5.21$\pm$0.24 & &   13 & 10/7  00:57:44 & 192.040 & 1834  & 0.217$\pm$0.015\\      
14 & 12/7 15:06:01 & 194.629 & 608  & 5 & 3.93$\pm$0.23 & &   14 & 11/7  07:16:13 & 193.303 & 2238  & 0.215$\pm$0.013\\      
15 & 13/7 09:57:13 & 195.415 & 576  & 3 & 3.93$\pm$0.29 & &   15 & 12/7  00:42:46 & 194.030 & 1915  & 0.221$\pm$0.014\\      
16 & 14/7 16:48:41 & 196.701 & 512  & 3 & 5.25$\pm$0.34 & &   16 & 13/7  03:51:19 & 195.161 & 1963  & 0.178$\pm$0.013\\
   &               &         &      &   &               & &   17 & 14/7  03:42:38 & 196.155 & 1901  & 0.231$\pm$0.014\\\hline
\end{tabular}

\medskip
Notes: $^a$Observation No., $^b$Middle of the observation (UT),
$^c$Where Jan 1$^{\rm st}$ 1996 at 00:00:00 is day 1.0 (UT),
$^d$Exposure (seconds) after data selection, $^e$No. of PCUs on during
the observation ,$^f$2.19--15.5\,keV band (spectral channels 2-36,
systematic effects of the background modeling {\em not} included),
when only three detectors were on the count rate has been corrected using
PCU area ratios
\end{minipage}
\end{table*}

The main instrument on {\sl RXTE} \cite{BRA93} is the proportional
counter array (PCA). With five xenon filled counters (PCU's), it has
an energy range of 2-60\,keV, an energy resolution of 18 per cent at
6\,keV, a large ($\sim$0.7m$^2$) effective area, and a circular field
of view (FWHM: 1$^{\rm o}$). Each PCU has three separate layers. Only
data from layer 1 has been used, as for observations of faint sources
most of the counts in layers 2 and 3 are caused by the background. Problems
with two of the PCUs meant they were switched off for some of the
observations (noted in Table~\ref{tab-xteobs}).

The {\sl RXTE} observations (see Table~\ref{tab-xteobs}) were made on
a roughly daily basis between 1996 June 27 to July 14 and have an
average exposure of about 650 seconds. Data reduction was done using
{\sl RXTE} specific {\sc ftool} programs. The Standard2 data were
filtered using internally applied good times (e.g. times of SAA
passage), as well as the criteria that elevation above the horizon of
greater than 10$^{\rm o}$ and offsource angle of less than 0.1$^{\rm
o}$, and then reduced to light curves and spectra using {\sc saextrct}.

The PCA cannot measure the background during an observation and so it
must be estimated. The accuracy of this estimation is very important
for faint sources such as 3C 279. {\sc pcabackest} (v1.4g with the q6
model) was used to produce models of the background due to particles
and cosmic X-rays, and spectra and light curves were then extracted
from these models in the same way as for the source observations. In
order to check the accuracy of the PCA background model, and to
provide a systematic error for the PCA datapoints, we have analysed
11 slew observations with individual exposures of greater than 400
seconds (average exposure is 671 seconds). The mean residual count
rate after subtraction of the background in the $\sim$2--15\,keV band
is -0.25 cts/sec (i.e. the model is an over estimate of the data on
average), with an intrinsic scatter of 0.44 cts/sec.

{\sl ROSAT} observed 3C 279 17 times between 1996 June 28 and
July 14 (see Table~\ref{tab-xteobs}) using the {\sl HRI}, an imager
with an energy range of 0.1--2\,keV, but no energy resolution. The
data were analysed using {\sc xselect}, part of the {\sc ftool}
software system. Counts were taken from a circular region 160 arcsec in
radius centred on the source, while the background was derived from an
annulus 160--320 arcsec from the source.

\section{Light Curves}
\begin{figure}
\begin{center}{
 \epsfxsize 1.0\hsize
 \leavevmode
 \epsffile{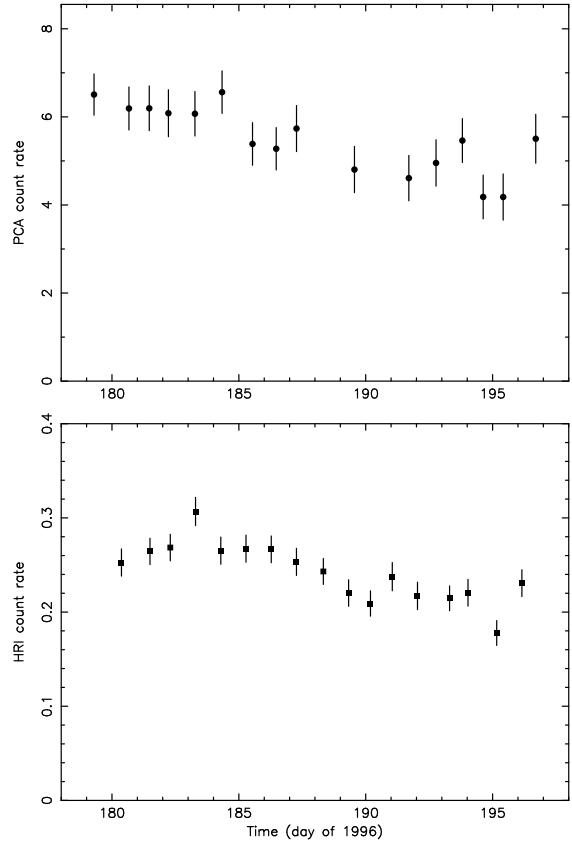}
}\end{center}
\caption{{\sl RXTE} (top) and {\sl ROSAT} light curves of 3C 279. The {\sl RXTE} points include the effects of the background modelling discussed in Section~\ref{sec-obs}}
\label{fig-xtelc}
\end{figure}

Visual examination shows similar behaviour in both the {\sl RXTE} and
{\sl HRI} light curves (Fig.~\ref{fig-xtelc}). Both light curves display
a slow, quasi-linear, decrease in intensity of approximately 30 per
cent.  A discrete cross-correlation does not provide any strong
constraints, owing to the largely featureless nature of the light curves,
and the two light curves are perfectly consistent with zero lag.

\section{Spectral analysis}

\subsection{Broad-band comparisons}

\begin{figure}
\begin{center}{
 \epsfxsize 1.00\hsize
 \leavevmode
 \epsffile[18 -586 774 -24]{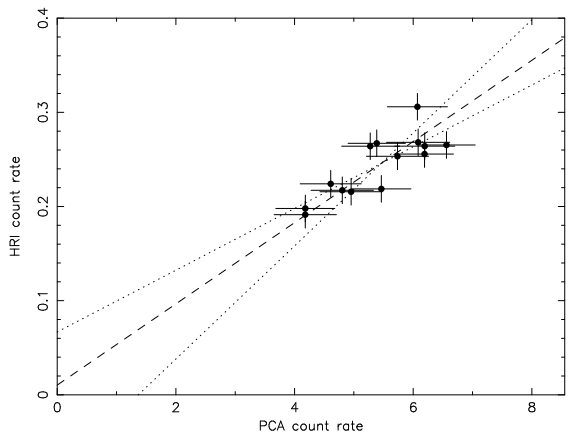}
}\end{center}
\caption{Shifted {\sl HRI} vs unshifted {\sl RXTE} count rate. The dashed line shows the best-fitting straight line model to the data. The dotted lines on the plots
show the 68 per cent upper and lower bounds for this model.}
\label{fig-hric2s}
\end{figure}

Before performing detailed spectral analysis, we have looked for any
correlations between the soft ({\sl ROSAT}) and hard ({\sl RXTE})
X-ray count rates.  The {\sl ROSAT} and {\sl RXTE} observations were
not simultaneous, so to compare the two light curves we have linearly
interpolated the {\sl ROSAT} data to the time of the {\sl RXTE}
observations.  The error for the shifted {\sl HRI} points is the mean
of the errors listed in Table~\ref{tab-xteobs}. Fig.~\ref{fig-hric2s}
shows the shifted {\sl ROSAT} and unshifted {\sl RXTE} count rates
plotted against one another. A simple straight line model is not a
good fit (reduced $\chi^2 \sim 2$). This may be a result of changes in
the relative amounts of hard and soft components or spectral changes
to a single component. Investigation shows that changes of $\pm7$ per
cent in the hardness ratio cannot be ruled out.  We have also, as a
check, interpolated the {\sl RXTE} data to the time of the {\sl ROSAT}
observations. The results are consistent with those (above and below)
for the opposite case.

Of interest with respect to the soft excess is the intercept on the
y-axis. If a single component spanned both the soft and hard X-ray
bands, then the intercept would be expected to be at zero. However, if
a soft excess were present then some residual soft count rate should
remain, as the hard band counts reduce to zero. The best-fitting models
intersect the y-axis near zero, suggestive of no soft excess. However,
the poorly sampled counts--counts plane and the large error bars give
a mean 99 per cent upper limit on the residual {\sl HRI} count rate of
0.15 counts s$^{-1}$. This is a sizable fraction of the observed
{\sl HRI} count rate, and only places a weak constraint on any soft
excess of $\leq$60 per cent of the soft X-ray count rate.

\subsection{Spectral fitting}
\label{sec-spec}
Spectral analysis was performed using the {\sc xspec} fitting package, with
the results shown in Table~\ref{tab-spec}. A power-law model with
Galactic absorption (see \ncite{ELV89}) gives an acceptable fit to
each {\sl RXTE} observation. All the best-fitting indices are consistent
with a single power law slope of $\alpha = 0.78\pm0.03$. This lies in
the range measured by {\sl ASCA} and {\it Ginga} \cite{TAS94et,LAW97}.

The same model was applied to the combined {\sl RXTE} and {\sl ROSAT}
data. We have treated the observations from each satellite as
simultaneous if they were made within six hours of each other. If this
was not the case, then the {\sl RXTE} observation has been fitted
separately with the two {\sl ROSAT} observations (pre-~and post-{\sl RXTE})
closest to it, with the results being averaged (with errors added in
quadrature). As with the {\sl RXTE} data alone, a single power law is
an acceptable fit in all cases. The errors on the derived spectral
indices are much reduced due to the extended energy range, but again
all the data are consistent with a single power law of $\alpha =
0.74\pm0.01$.  This is consistent with that derived from the {\sl
RXTE} data alone, and means that the soft X-ray data are consistent
with an extension of the hard X-ray power law into the {\sl HRI}
energy band.

\begin{figure}
\begin{center}{
 \epsfxsize 1.07\hsize
 \leavevmode
 \epsffile[18 -586 774 -24]{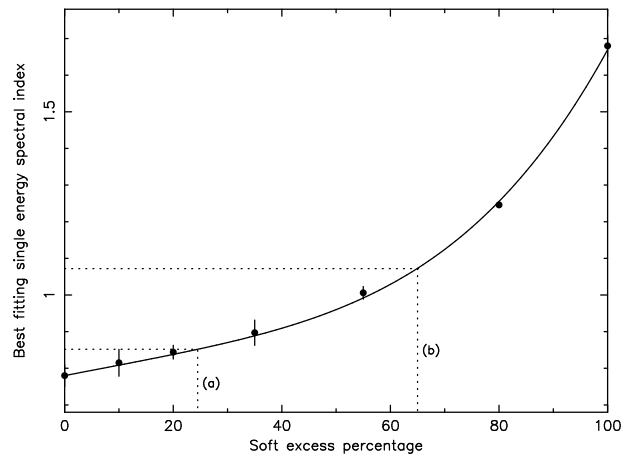}
}\end{center}
\caption{Plot of spectral index vs soft excess percentage derived from modelling. The data points represent the mean
of the best-fitting indices for a particular strength of soft excess,
the error bars are the standard deviation of these best fits (there
are 20 model spectra per data point). Line (a) shows the upper limit
for 3C 279 derived from the spectral fitting, while line (b) shows the
expected spectral index if the soft excess were as prominent as that
in 3C 273}
\label{fig-model}
\end{figure}

However, as the {\sl HRI} has no spectral capability a soft excess
could still be hidden. To investigate this we produced a number of
fake source spectra with two power law components: a hard component
that is the average of the fits to the {\sl RXTE} data alone
($\alpha=0.78$ with a flux at 1\,keV of $3\times10^{-3}$
photons\,cm$^{-2}$\,s$^{-1}$\,keV$^{-1}$), and a soft component
($\alpha=1.68$, the average value for the soft excess of 3C 273, Leach
et al. 1995) with a varying normalisation. A single power law has then
been fitted to these spectra (acceptably in each case) and the
best-fitting spectral index measured. The results of the fitting are
shown in Fig.~\ref{fig-model}. As expected, a larger fraction of soft
excess results in a steeper single power law index.

The best-fitting spectral index to the combined {\sl ROSAT} and {\sl
RXTE} data is $\alpha=0.74\pm0.04$ (where the error is the standard
deviation of the best fit values), and the 99 per cent, or
2.6\,$\sigma$ upper limit is 0.84. This gives an upper limit to the
luminosity of a soft excess of $\sim25$ per cent of the total
luminosity in the 0.1--2.0\,keV band, or less than $3\times10^{38}$\,W
(${\rm q}_0=0.5, H_0=50$\,km\,s$^{-1}$\,Mpc$^{-1}$).

For comparison we have calculated (from Leach et al. 1995) that on
average 73 per cent of the 0.1--2.0\,keV luminosity of 3C 273 is from
the soft excess. This is much larger than that seen in 3C
279. Red-shifting the spectrum of 3C 273 to the distance of 3C 279
only reduces the soft excess percentage to 65 per cent. If this
percentage were actually present in 3C 279, we would expect an average
spectral index of near $\alpha=1.1$, a value some 8$\sigma$ away from
the observed value. This clearly shows that any soft excess in 3C 279
is intrinsically weaker relative to the hard X-ray power law
than is seen in 3C 273.

\begin{table}
\caption{Spectral fitting results}
\label{tab-spec}
\begin{tabular}{@{}l@{$\;\;\:\,$}ccccccc@{}}\hline
N$^a_{\rm X}$ & $\alpha^b$ & $\chi^{2c}_\nu$ &  N$^d_{\rm R}$ & $\alpha^e$ & $\chi^{2f}_\nu$ & S$^g$ & S$^h$ \\\hline
1  & 0.82$\pm$0.08          &  0.5  & 1     & 0.69$\pm$0.03 & 0.6 & 12.8           & 8.3 \\
2  & 0.68$\pm$0.11          &  1.1  & 1,2   & 0.70$\pm$0.05 & 1.0 & 12.3           & 8.1 \\
3  & 0.58$\pm$0.13          &  0.6  & 2     & 0.72$\pm$0.04 & 0.6 & 12.1           & 8.2 \\
4  & 0.80$^{+0.17}_{-0.16}$  & 0.8  & 3     & 0.75$\pm$0.04 & 0.8 & 11.8           & 8.5 \\
5  & 0.68$^{+0.13}_{-0.12}$  & 0.9  & 4     & 0.80$\pm$0.04 & 0.9 & 12.1           & 9.5 \\
6  & 0.83$\pm$0.10          &  0.8  & 5     & 0.70$\pm$0.03 & 0.8 & 13.0           & 8.6 \\
7  & 0.99$^{+0.13}_{-0.12}$  & 0.5  & 6,7   & 0.80$\pm$0.05 & 0.6 & 10.8           & 8.5 \\
8  & 1.02$\pm$0.12           & 0.6  & 7     & 0.81$\pm$0.04 & 0.7 & 10.7           & 8.6 \\
9  & 0.66$^{+0.15}_{-0.14}$  & 0.7  & 8     & 0.72$\pm$0.04 & 0.7 & 11.5           & 7.9 \\
10 & 0.45$\pm$0.18          & 0.2  & 10    & 0.75$\pm$0.05 & 0.3 & \phantom{1}9.5 & 6.8 \\
11 & 0.68$^{+0.20}_{-0.19}$ & 0.9  & 12,13 & 0.80$\pm$0.07 & 0.9 & \phantom{1}9.0 & 7.1 \\
12 & 0.65$^{+0.19}_{-0.18}$ & 0.4  & 13,14 & 0.73$\pm$0.07 & 0.4 & \phantom{1}9.8 & 6.8 \\
13 & 0.96$\pm$0.14         &  0.7  & 15    & 0.70$\pm$0.04 & 0.8 & 10.9           & 7.2 \\
14 & 0.73$\pm$0.19         &  0.7  & 15,16 & 0.77$\pm$0.07 & 0.6 & \phantom{1}8.3 & 6.2 \\
15 & 0.82$^{+0.22}_{-0.21}$ & 0.6  & 16,17 & 0.77$\pm$0.08 & 0.6 & \phantom{1}8.5 & 6.4 \\
16 & 0.76$^{+0.20}_{-0.19}$ & 0.5  & 17    & 0.71$\pm$0.05 & 0.5 & 10.8           & 7.3 \\\hline
\end{tabular}

\medskip
Notes: $^a${\sl RXTE} observation number, $^b$best-fitting power law
slope to {\sl RXTE} data only with 68 per cent error, $^c$value for
{\sl RXTE} fitting only, $^d${\sl ROSAT} observation numbers used in
combined fits with the {\sl RXTE} data, $^e$best-fitting power law slope
to the combined {\sl RXTE} and {\sl ROSAT} data (see
Section~\ref{sec-spec}), $^f$value for combined {\sl ROSAT} and {\sl RXTE}
fits, $^g$2--10\,keV model flux, $\times10^{-15}$\,W\,m$^{-2}$,
$^h$0.1--2\,keV model flux, $\times10^{-15}$\,W\,m$^{-2}$
\end{table}

\section{Conclusions}

We have presented contemporary {\sl RXTE} (2--15\,keV) and {\sl ROSAT}
(0.1--2\,keV) observations of the quasar 3C 279 at daily intervals
over a three week period. During this time the source exhibited only
low amplitude variability, decreasing in flux by $\sim$30 per cent. The
largely featureless nature of the light curves prevents us from
determining any lag between the {\sl RXTE} and {\sl ROSAT} light curves
which are quite consistent with zero time delay. We find no evidence
for any spectral variability either within the {\sl RXTE}
observations or in the combined {\sl RXTE} and {\sl ROSAT}
observations.

There is also no significant evidence that a soft excess is present,
although modelling cannot rule out a soft excess which emits 25 per
cent of the 0.1--2.0\,keV X-ray luminosity. Comparison with 3C 273
shows that any soft excess is relatively weaker in 3C 279 than in 3C
273 (although the upper limit to the soft excess in 3C 279 in terms of
absolute luminosity is some three times larger than that of 3C
273). This relative weakness could be caused by a number of phenomena, such
as a more luminous source having a more massive black hole and cooler
disc, but it is more likely that 3C 279 is more jet dominated than 3C 273,
perhaps due to a higher beaming factor or more acute viewing angle.

\end{document}